\documentclass[reprint,amsmath,amssymb,aps,pra,longbibliography,nofootinbib]{revtex4-2}
\usepackage{algorithmic}
\usepackage{graphicx}
\usepackage{textcomp}
\usepackage{lipsum,mathtools,cuted}
\usepackage{float}
\usepackage{epsfig}
\usepackage{epstopdf}
\usepackage{subfigure,color}
\usepackage{dcolumn}
\usepackage{accents}
\usepackage{multirow}
\usepackage[usestackEOL]{stackengine}
\usepackage{tabularx}
\usepackage[normalem]{ulem}
\usepackage{makecell}
\usepackage{dcolumn}
\usepackage{bm}
\usepackage{booktabs} 
\usepackage{comment}
\usepackage{times}
\usepackage{amsfonts}
\usepackage{esvect}
\usepackage{harpoon}
\usepackage{amsmath,accents}
\usepackage{subfigure,color}
\usepackage{graphicx}
\usepackage{dcolumn}
\usepackage{bm}
\usepackage{comment}
\usepackage[mathlines]{lineno}
\usepackage{xcolor}
\usepackage{environ} 
\newif\ifwc \wctrue    

\begin{document}

\title{Non-Foster Photonic Time Crystals}
\author{Z.~Li$^{1}$}
\author{M.~S.~Mirmoosa$^{2}$} 
\author{S.~Hrabar$^{3}$} 
\author{V.~Asadchy$^{4}$} 
\author{X.~Wang$^{5}$} 
 
\affiliation{$^{1}$College of Information and Communication Engineering, Harbin Engineering University, China\\$^{2}$Department of Physics and Mathematics, University of Eastern Finland, P.O.~Box~111, FI-80101 Joensuu, Finland\\$^{3}$University of Zagreb Faculty of Electrical Engineering and Computing, Unska 3, HR-10000 Zagreb, Croatia\\$^{4}$Department of Electronics and Nanoengineering, Aalto University, P.O.~Box~15500, FI-00076~Aalto, Finland\\$^{5}$College of Physics and Optoelectronic Engineering, Harbin Engineering University, China}


\begin{abstract}

Photonic time crystals (PhTCs) are spatially uniform media whose material parameters vary periodically in time, opening momentum bandgaps within which the fields of  electromagnetic modes can grow exponentially in time. To date, PhTCs have utilized only passive, lossless materials with ``positive" dispersion (Foster materials), and a theoretical framework addressing active materials with ``negative" dispersion (non-Foster materials) in PhTCs and their associated physical properties remains undeveloped. 
 Here, we explore the two classes
of isotropic PhTCs with embedded non-Foster inclusions: a bulk medium with periodically modulated negative permittivity, and a metasurface whose surface capacitance alternates between positive and negative values. Employing an analytical transfer-matrix formulation, we demonstrate that non-Foster permittivity modulation not only broadens momentum bandgaps without bounds but also provides a gain rate that increases linearly with momentum. Remarkably, the proposed isotropic PhTCs support exponential amplification down to zero frequency—a regime inaccessible in conventional isotropic PhTCs. These results open new avenues for ultra-broadband wave control, high-gain signal processing, and energy-harvesting devices that leverage the unique dispersion of active, time-modulated circuitry.
    
\end{abstract}

\maketitle





    


\begin{figure*}
    \centering
    \includegraphics[width=1\linewidth]{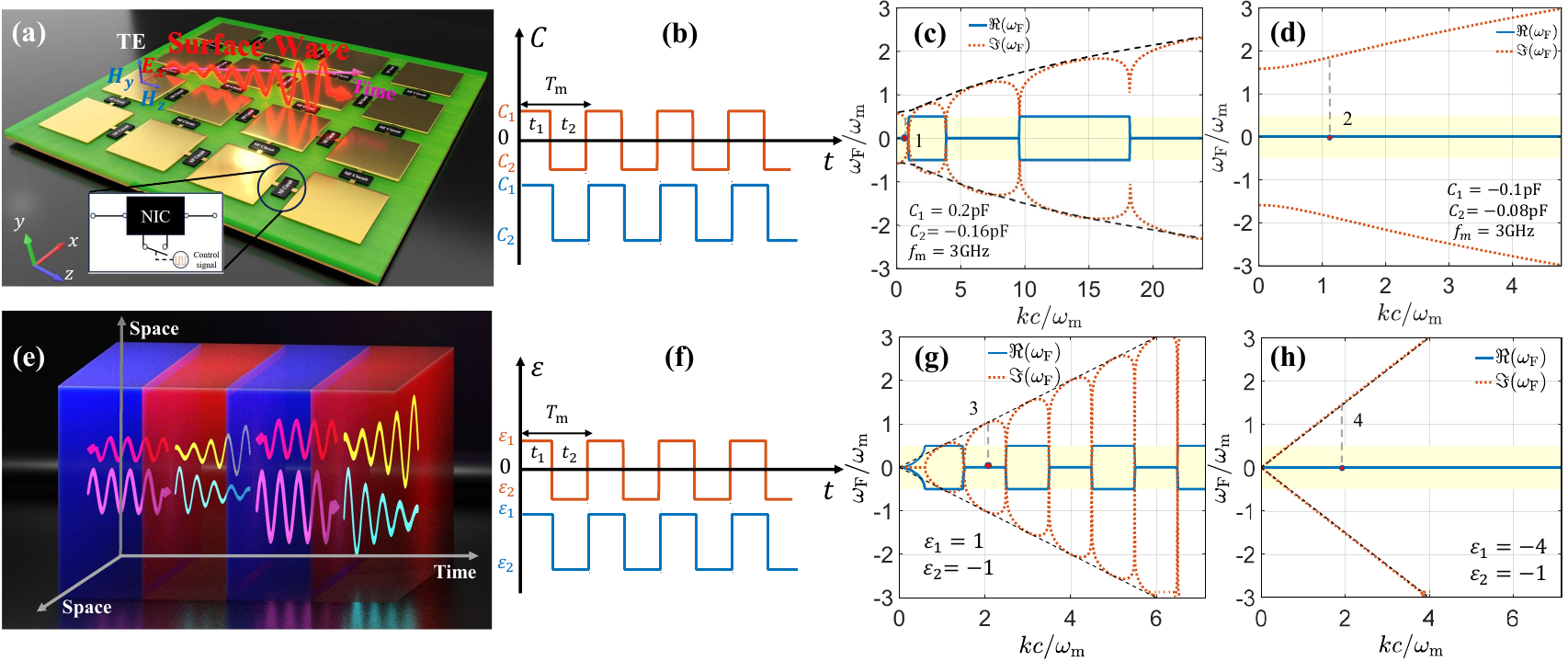}
    \caption{(a), (e) Geometry structure of the two types of PhTCs studied in this work. (b),(f) The corresponding modulation profiles of their material properties (the red line represents the modulation between positive and negative capacitance or permittivity values, whereas the blue line represents the modulation between two negative capacitance or permittivity states). (c), (d), (g), (h) The corresponding band structures. (a) Metasurface-based PhTC. The zoomed-in view of the non-Foster circuit (NF circuit) represents the time-varying negative capacitor integrated between two adjacent unit cells. The negative capacitor is created by the negative impedance converter (NIC) (internally loaded with a capacitor), whose input capacitance is controlled  by an external square wave signal (please see Fig.~\ref{fig:1}(b) in Supplementary materials). (e) Bulk media PhTC, whose relative permittivity is modulated by a periodic square function shown in (f). (c), (g) The band structures for the scenarios where capacitance and permittivity, respectively, are modulated between the positive and negative values. (d), (h) Same as (c), (g), but for the case when capacitance and permittivity are modulated between 2 negative values. The real eigenfrequencies belonging to the first Brillouin zone only are plotted, as indicated by the yellow region. The black dashed line represents the envelope function of the imaginary part of the eigenfrequency. (c) and (d) depict the band structure of the metasurface, while (g) and (h) show the band structure of bulk media. In the calculations, both \( t_1 \) and \( t_2 \) are set to \( T_\mathrm{m}/2 \). The red points and gray dashed lines marked with numbers from ``1'' to ``4'' indicate points selected for calculating the corresponding harmonic distributions and electric field distributions.}
    \label{fig:1}
\end{figure*}


Materials with temporally modulated optical properties have opened promising frontiers in controlling interactions of classical~\cite{galiffi2022photonics, engheta2023four, sounas2017non, ramaccia2020electromagnetic} and quantum~\cite{mendoncca2000quantum,mirmoosa2023quantum} light fields with matter. Of particular note are photonic time crystals (PhTCs), which exhibit the remarkable characteristics such as momentum bandgaps~\cite{asgari2024theory}. This unique feature enables complex eigenfrequencies, leading to the exponential amplification of field amplitudes over time~\cite{asgari2024theory}. Research on PhTCs has advanced rapidly, with theoretical investigations evolving from simple periodic refractive index modulations~\cite{zurita2009reflection,martinez2016temporal,CT_Chan,kodama2025magnetic} to complex structured designs incorporating dispersion~\cite{wang2025expanding,ozlu2025floquet,feng2024temporal}, anisotropy~\cite{li2023stationary}, and bianisotropy~\cite{koufidis2024electromagnetic}, complemented by growing experimental efforts~\cite{reyes2016electromagnetic,park2022revealing,wang2023metasurface}. This progress has unveiled a diverse landscape of potential applications, including thresholdless lasing~\cite{lyubarov2022amplified}, enhanced and controlled emission~\cite{dikopoltsev2022light,gao2024free,zhu2025smith}, stationary charge radiation~\cite{li2023stationary}, polarization-selective amplification~\cite{koufidis2024electromagnetic}, among others. 

Despite this significant progress in the field, the prior studies on PhTCs remain fundamentally constrained by their dependence on material parameters whose electromagnetic responses are governed by energy-dispersion constraints \cite{landau_Electrodynamics_1984} or, in circuit-theory language, by Foster's reactance theorem \cite{foster_Reactance_1924}. This theorem establishes that the input reactance ($X$) or input susceptance ($B$) of every passive, lossless network increases monotonically with frequency ($\partial{X}/\partial\omega>0$), ($\partial{B}/\partial\omega>0$) \cite{foster_Reactance_1924,collin}. Overcoming this limitation requires the introduction of reactive elements with negative values, so-called non-Foster elements, which enable the realization of broadband responses~\cite{hrabar2018first,mou2016design} that are otherwise unattainable with conventional passive components. In material terms, this corresponds to engineering lossless media with constant negative permittivity or negative permeability. Against this backdrop, the investigation of PhTCs utilizing non-Foster materials or structures emerges as a promising and innovative direction, which may offer novel and intriguing optical functionalities. Notably this effort is different from the temporal modulation of active materials with lossy and/or gainy characteristics~\cite{li2021temporal}. 

In this Letter, we scrutinize this unexplored potential, i.e.,~PhTCs employing non-Foster materials, and, in particular, we concentrate on two distinct systems. The first is an impenetrable metasurface composed of a dense array of patches connected with capacitors whose capacitance is periodically switched between positive and negative values. The second system is a bulk medium in which the permittivity is periodically switched either between positive and negative values or between two different negative values. The resulting band structures reveal interesting phenomena such as the amplification of zero-frequency (DC) waves and exotic momentum bandgaps with the imaginary eigenfrequencies linearly growing with the wavenumber. These findings provide new insights into the interplay between temporal periodicity and active non-Foster material properties, paving the way for advanced applications in broadband signal processing, energy harvesting, and beyond.

We commence the investigation by formally introducing the approach to calculate the band structure of a periodically modulated metasurface. Consider an impenetrable metasurface composed of subwavelength metallic patches and located in the $xz$ plane in vacuum, see Fig.~\ref{fig:1}(a). As a result of the capacitive coupling between the neighbouring patches, this metastructure is effectively homogenized and represented by a macroscopic parameter called effective capacitance. 
Suppose that a transverse electric (TE) polarized wave propagates along this time-varying metasurface. The reason for selecting this polarization comes from the fact that, for a static capacitive sheet, when the effective capacitance is positive, a surface wave with TE polarization and real-valued frequency is supported according to Maxwell's equations. On the other hand, we prove in the following that such a polarization is also supported if the value of the effective capacitance is negative, although the frequency has a different characteristic. In Fig.~\ref{fig:1} (a), we assume that the metasurface is static (i.e., assume that the integrated capacitors are immutable) and $+z$ is the propagation direction. Hence, we write the transverse electric and magnetic field components ($E_x$ and $H_z$) related to the TE polarization as: $E_x=E_0 e^{-\alpha y} e^{-j \beta z} e^{j\omega t}$ and $H_z=[-\alpha E_0/(j\omega \mu_0)]e^{-\alpha y}e^{-j \beta z} e^{j\omega t}$, respectively. Here, $E_0$ is the complex amplitude, $\alpha$ is a positive real-valued attenuation coefficient representing the decay rate along the $+y$ direction, and $\beta$ is the propagation constant along the $+z$ direction. Since the metasurface is surrounded by free space, we apply the Helmholtz equation to $E_x$ and obtain $\alpha^2-\beta^2+{\omega^2}/{c^2}=0$, where $c$ is the speed of light. Besides, we impose the impedance boundary condition at $y=0$, which is expressed as $H_z=j\omega CE_x$, $j\omega C$, being a surface admittance caused by presence of the total effective capacitance $C$.  By combining these results, one obtains the corresponding dispersion relation and another important relation about the attenuation coefficient given by~\cite{ma2019parallel} 
\begin{equation}
\beta^2=\frac{\omega^2}{c^2}\Bigg(1+\frac{\mu_0}{\varepsilon_0} \omega^2 C^2\Bigg),\quad\alpha=\omega^2\mu_0C.
\label{eq:4Four}
\end{equation} 
Remember that we focused on surface waves; that is, $\alpha$ must be a positive real value. There are two scenarios for $C$ and $\omega$ according to Eq.~(\ref{eq:4Four}) to satisfy this criterion. One is that $C>0$, and therefore, $\omega$ must be a real value. The other is $C<0$, which corresponds to a non-Foster metasurface, and $\omega$ must be a purely imaginary value (i.e.,~$\omega = \pm jg$, in which $g$ is a real value). For this specific scenario, the surface wave allowed with angular frequency $\omega = \pm jg$ is, in fact, a frozen wave because of the absence of the real part of the frequency. Furthermore, the amplitudes of this surface wave exhibit exponential decay or growth in time, following the behaviour $e^{\pm gt}$. Meanwhile, except for positive $\alpha$, $\beta^2$ must also be positive to avoid a complex propagation constant. To satisfy this condition, Eq.~\eqref{eq:4Four} states that there is a second requirement for the angular frequency, which is $\vert\omega\vert^2>\varepsilon_0/(\mu_0C^2)$. Based on the above discussion, we see that the TE polarization is supported regardless of the sign of the effective capacitance, making the postulate of its existence appropriate for the proposed time-varying metasurface. 

To continue, we examine the case where the effective capacitance of the metasurface is modulated by an ideal periodic square function which is expressed as $C(t)=C_2$ when $(n-1) T_\mathrm{m}<t<(n-1) T_\mathrm{m}+t_2$ and $C(t)=C_1$ when $n T_\mathrm{m}-t_1<t<n T_\mathrm{m}$, where $n \in \mathbb{Z}$ (see Fig.~\ref{fig:1}(b)). Notice that the theory derived below is applicable independent of the sign of $C_1$ and $C_2$ capacitance values. For such a periodically modulated metasurface, the transfer-matrix method~\cite{asgari2024theory} is employed to derive the band structure.

It is well established that when the electromagnetic property of a material is abruptly changed in time, while remaining uniform in space, the wave momentum remains unchanged even if the wave is frozen in space after switching as shown in~\cite{pacheco2025holding}. However, the frequency becomes adjusted to satisfy the new dispersion relation after the jump \cite{asgari2024theory}. Assume the transverse propagation constant and frequency to be (\(\beta\), \(\omega_1\)) corresponding to time segment 1, and (\(\beta\), \(\omega_2\)) to time segment 2. Since $\beta$ is conserved, $\omega_1$ and $\omega_2$ are related through the dispersion relation (\ref{eq:4Four}): $\omega_1^2(1+\frac{\mu_0}{\varepsilon_0}\omega_1^2 C_1^2)=\omega_2^2(1+\frac{\mu_0}{\varepsilon_0}\omega_2^2 C_2^2)$. One point that should be clarified here is that the momentum conservation equation always gives one positive and one negative solution for $\omega_2^2$. However, to keep $\alpha$ positive, the positive solution must be discarded (see Eq.~\eqref{eq:4Four}). Thus, we only have one valid solution for  $\omega_2$. Accordingly, the electric field can then be expressed for each time segment within the \(n\)-th period of the time evolution as a summation of either the backward and forward waves for real frequency or the time-growing and time-decaying waves for purely imaginary frequency. Then, by using the temporal continuity conditions at the moments \((n-1)T_\mathrm{m}\) and \(nT_\mathrm{m} - t_1\), the transfer matrix can be obtained. Next, in order to derive the band structure, Floquet theorem is utilized, and an eigenvalue problem is finally formulated, which needs to be solved to calculate the band structure. The detailed derivation can be found in Supplementary Materials, Section 1. Here, we just show the final eigenvalue equation:
\begin{equation}
\Big(\mathbf{A}-e^{-j\omega_{\rm F} T_m} \mathbf{I}\Big)(a_{n},b_{n})^T
=0, 
\label{eq: 14}
\end{equation}
where $\mathbf{I}$ denotes an identity matrix, $\mathbf{A}$ is a square matrix, and $\omega_{\rm F}$ refers to the Floquet frequency. 

Similarly, for a time-varying lossless bulk medium with a square-function-modulated permittivity as in Figs.~\ref{fig:1}(e)-(f), the band diagram can be derived using the same method. Hence, we elaborate on it in Supplementary Material in Section~1.



Now, we discuss the band structure calculated using the theory developed above, which yields some interesting physical phenomena. Several examples demonstrate the band structure of the capacitive metasurface and the bulk medium modulating in both positive-negative and negative-negative cases, Fig.~\ref{fig:1}(c), (d), (g), (h). These findings are derived assuming an ideal, non-dispersive model of negative impedance~\cite{hrabar2018first, hrabar_Negative_2011, hrabar_Ultrabroadband_2013, okorn_Physically_2017}. As we discuss in detail in the Supplemental Material in Section 2, realistic non-Foster circuits can be engineered to approximate very well this ideal behavior over a significant bandwidth \cite{buiantuev_Physically_2022}. Thus, these novel phenomena are certainly experimentally accessible and highly relevant for practical implementations. Notably, a similar situation had occurred in the past, in the case of broadband non-Foster metamaterials that were first predicted theoretically with an assumption of ideal elements \cite{tretyakov_Metamaterials_2001} and later successfully experimentally demonstrated by several groups in various engineering systems,  as reviewed in \cite{hrabar2018first}. Therefore, we focus on the ideal model here to elucidate the core physics in the clearest possible setting.


First, we analyze the scenario when the metasurface capacitance or permittivity of the bulk material are modulated between the positive and negative values ($C_1>0$ and $C_2<0$ or $\varepsilon_1>0$ and $\varepsilon_2<0$). As shown in Fig.~\ref{fig:1}(c) and (g), over the entire range of $\beta$ values, the real part of the eigenfrequency within the momentum band gaps can take only values of either \( n\omega_{\rm m} \) or \( (2n+1)\omega_{\rm m}/2 \), where \( \omega_{\rm m} \) is the modulation frequency. In the figures, we show the real eigenfrequencies belonging to the first Brillouin zone. Meanwhile, the imaginary part forms a unique pattern of ellipses arranged sequentially along the wavenumber axis, indicating a nearly infinite bandgap. This unique band structure is quite similar to the one found in photonic crystal made by periodically arranged positive and negative index materials~\cite{li2003photonic}. Interestingly, in contrast to conventional PhTCs, where the imaginary eigenfrequency within the momentum band gap usually forms an elliptical shape, in the case of non-Foster PhTCs, the imaginary eigenfrequency increases with \( \beta \) and only drops sharply at the band gap edges. It is worth noting that for a given $\beta$, all frequency harmonics separated by $\omega_{\rm m}$ (i.e., $\Re(\omega_{\rm F})+ n\omega_{\rm m}$) share the same imaginary eigenfrequency $\Im(\omega_{\rm F})$.
The envelope of the imaginary eigenfrequency exhibits a continuous growth as the momentum increases. Interestingly, we find that the slope of the envelope of the imaginary eigenfrequency for bulk media can be approximately expressed as:
\begin{equation}
    \Im(\omega_{\rm F}) \approx \frac{c\left(\sqrt{\varepsilon_1/\left|\varepsilon_2\right|}+1\right)}{2\sqrt{\varepsilon_1}}k, \label{eq: 15}
\end{equation}
where \( c \) is the speed of light in vacuum. Here, this approximation is accurate when \(\varepsilon_1\) and \(|\varepsilon_2|\) are close to each other (with a difference of no more than approximately tenfold). 

The described band diagrams imply that if a pulse propagates inside a non-Foster PhTC,   all its momentum components will be amplified, with higher momentum components experiencing a larger amplification rate. Consequently, the high-\(k\) components will eventually dominate inside the pulse. This exotic behavior can be leveraged to effectively enhance the near fields of dipole radiation.

We also notice from the band structures in Fig.~\ref{fig:1} that the zero-momentum mode (i.e., the wavenumber is zero) exhibits distinct behavior for the bulk media PhTC and metasurface-based PhTC. In the case of the bulk PhTC, the imaginary eigenfrequency vanishes at $k=0$, whereas for the metasurface-based PhTC, it remains non-zero. This difference stems from the fundamental distinction in the mode supported by the bulk media and the metasurface. Bulk media support only propagating plane-wave modes. Therefore, when the wavenumber is zero, the dispersion relation $\omega^2 \varepsilon \mu=k^2$ enforces $\omega=0$, regardless of whether the permittivity is positive or negative. In contrast, a metasurface supports surface wave mode rather than plane wave mode. We can see from Eq.~(\ref{eq:4Four}) that when wavenumber is zero, the eigenfrequency can either be zero or a purely imaginary value. When the system is passive, only the zero-frequency solution is allowed. However, when the system is active, the purely imaginary eigenfrequency can exist. As a result, the zero-momentum mode in the metasurface-based PhTC exhibits a non-zero imaginary component in its eigenfrequency.

Moreover, it is noteworthy that the amplification rate can be significantly larger than that of conventional PhTCs for the same value of the modulation strength. Notably, \( |\Im(\omega_{\rm F})| \) can exceed \( |\Re(\omega_{\rm F})| \) for high-\( k \) values, see Fig.~\ref{fig:1}(c) and (g). This enhanced amplification rate is attributed to the introduction of non-Foster elements and holds great potential for applications in efficient optical communication, high-sensitivity optical sensing, and high-gain laser systems.

Next, we explore the scenario when the metasurface capacitance or the permittivity of the bulk material are modulated between two negative values ($\varepsilon_1<0$ and $\varepsilon_2<0$ or $C_1<0$ and $C_2<0$). As shown in Figs.~\ref{fig:1}(d) and (h), the real eigenfrequency can only take values of \( n\omega_{\rm m} \), while the imaginary part behaves like a continuously growing function of \( \beta \) (see orange line in Figs.~\ref{fig:1}(d) and (h)), exhibiting an unlimited amplification rate as the momentum increases. These intriguing features have not been observed in conventional PhTCs without non-Foster elements. In conventional PhTCs, the amplification rate, determined by the imaginary part of the eigenfrequency, is always limited, and the size of the momentum band gap is also constrained. Although a recent study~\cite{wang2025expanding} proposed a resonant structure to achieve an infinitely large band gap, the amplification rate within it remained inherently limited.

Different from positive-positive case, here, the imaginary eigenfrequency directly forms a tilted line with a positive slope in the band structure of bulk media. Similarly, this tilted line can be approximated as the form of Eq.~(\ref{eq: 15}). It also suggests that the imaginary eigenfrequencies of the time-varying medium can be predicted by a static medium with an effective negative permittivity of $\varepsilon_{\rm eff}= -\sqrt{2\sqrt{\varepsilon_1 \varepsilon_2}/(\sqrt{\left|\varepsilon_1 \right|}+\sqrt{\left|\varepsilon_2 \right|})}$. 

For the metasurface-based PhTC (see Fig.~\ref{fig:1}(c) and (d)), while not exactly linear growth of the amplification rate is observed, we still find the same trend where the amplification rate continuously increases with momentum. The features of the band structure of both metasurface and bulk media are quite similar to each other, indicating that these peculiar band structures are attributed to the non-Foster nature of both considered material systems.



Next, we show another characteristic of the non-Foster PhTC. In previous studies, it was believed that isotropic PhTCs cannot provide DC-wave amplification, and that temporal modulation of material anisotropy is necessary~\cite{li2023stationary}. However, as we demonstrate below, the non-Foster nature of our proposed PhTCs can provide an alternative path to achieve amplification of DC signals in isotropic systems. 



We arbitrarily select a value of \( k \) or \( \beta \), where the real part of the fundamental eigenfrequency is zero, while the imaginary part is non-zero, as indicated by the points marked from ``1'' to ``4'' in Fig.~\ref{fig:1} (c), (d), (g), (h). We then calculate the amplitudes distribution for different harmonics corresponding to these four eigenfrequencies (see more details in Supplementary Material, section 3), as shown in Fig.~\ref{fig:3}(a)-(d).
The normalized harmonic distribution in all four cases clearly shows a strong DC component ($n=0$) for which the field is not oscillating while growing in time due to the momentum bandgap. Meanwhile, we find that for positive-negative modulated bulk media, the amplitude of the DC wave, i.e., the $n=0$ order, is not always such large and is proportional to the ratio of the negative permittivity and positive permittivity. It means that the larger the negative permittivity, the stronger the DC field. This phenomenon comes from the ability to amplify DC signals of ideal non-Foster elements or any non-Foster elements that allow negative impedance to occur at zero frequency, see Supplementary Material in section 2.

\begin{figure}[t!]
    \centering
    \includegraphics[width=1\linewidth]{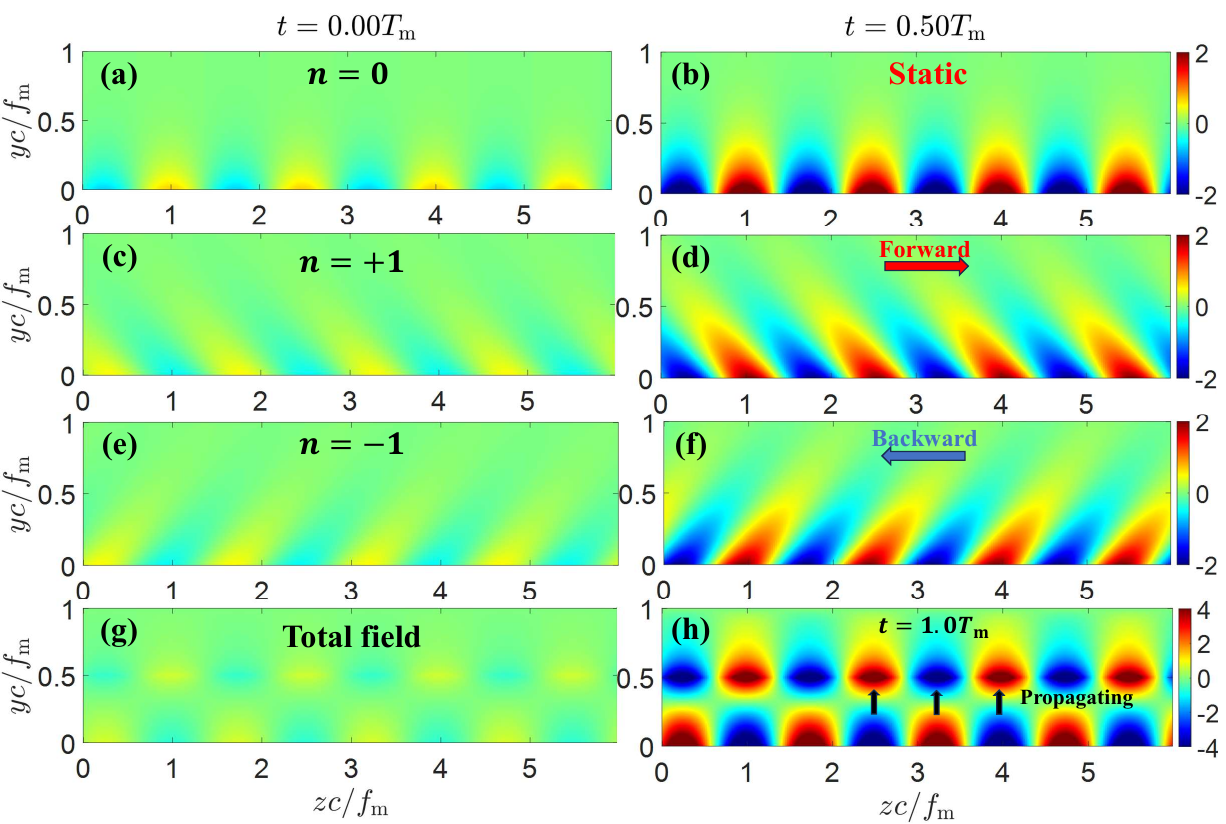}
    \caption{Analytical solution of the electric field distribution (time snapshots) over the metasurface-based PhTC evaluated for the eigenfrequency depicted by point ``3'' in Fig.~\ref{fig:1}(g). The units in the colorbar are V/m.   (a)-(b) The field evolution of the $n=0$ harmonic, i.e. DC harmonic with $\Re(\omega_{\rm F}) =0 $.
    (c)-(f) The field evolutions of the $n=\pm 1$ harmonics with $\Re(\omega_{\rm F}) =\pm \omega_{\rm m} $, respectively. (g)-(h) The total electric field including 50 frequency harmonics. The plots in the left panel show the field distribution at $t=0$, while (b), (d), (f) show the field at $t=\frac{T_{\rm m}}{2}$ and (h) at $t=T_{\rm m}$.}
    \label{fig:a}
\end{figure}

To more intuitively demonstrate the wave behavior within the momentum band gap, we calculate the electric field distribution for the eigenfrequency denoted by point ``1'' for the metasurface scenario in Fig.~\ref{fig:1}(c). Figures~\ref{fig:a}(a)--(f) depict these field distributions for the three dominant harmonics ($n=0$ and $n=\pm 1$) at two different time moments $t=0$ and $t=T_{\rm m}/2$. The field of the $n=0$ harmonic reveals a non-oscillating field pattern whose amplitude is growing in time. 

Interestingly, the fields of the $n=\pm 1$ harmonics correspond not to those of surface waves but propagating waves. Indeed, since we consider a wave with propagation constant $\beta$ within the momentum band gap, the eigenfrequency is consequently complex for the $n=\pm 1$ harmonics, and, therefore, the decay factor $\alpha$ in Eq.~(\ref{eq:4Four}) is no longer a real value but a complex value with positive real part and positive imaginary part (negative one should be discarded). Thus, the $n=\pm 1$ harmonics propagate and decay in a vertical direction and form a standing wave in the horizontal direction. One can see from Figs.~\ref{fig:a}(g) and (h) that the field is localized in space due to zero group velocity in the horizontal direction and amplified in time. Due to the presence of propagating wave harmonics (with propagation direction indicated by black arrows), the waves exhibit spatial decay in the vertical direction. However, their temporal amplification leads to additional enhancement, effectively extending the propagation distance.
 This phenomenon could be helpful in near-field sensing.


\begin{figure}[t!]
    \centering
    \includegraphics[width=1\linewidth]{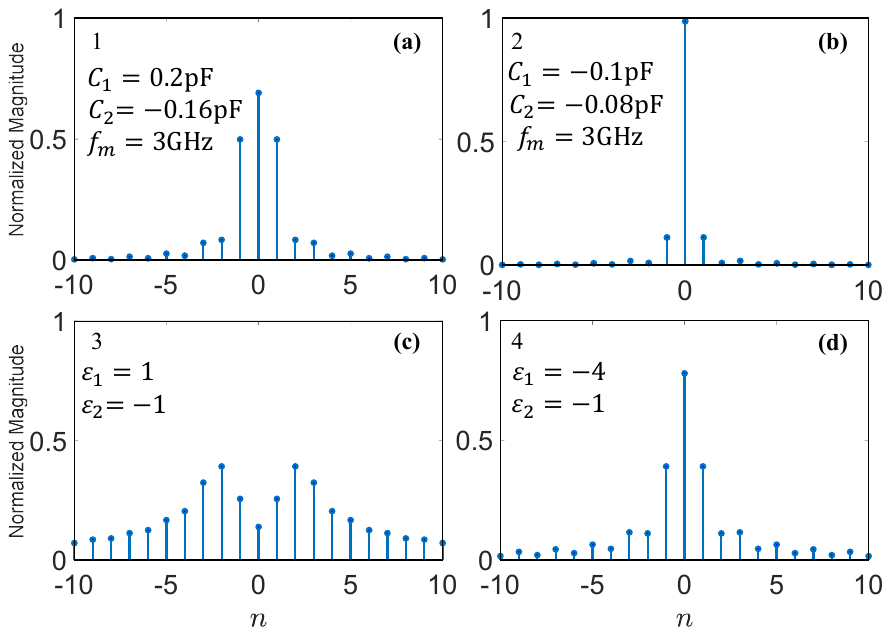}
    \caption{Amplitude distribution of the Floquet harmonics (normalized by energy) of the electric field for a given eigenfrequency of the PhTC based on (a)--(b) the metasurface with time-varying capacitance and (c)--(d) bulk medium with time-varying permittivity. 
    The numbers in the upper left corners of (a)-(d) correspond to the red points marked with the same numbers in Fig.~\ref{fig:1}, depicting the corresponding eigenfrequency.}
    \label{fig:3}
\end{figure} 

In this Letter, we have introduced the concept of non-Foster PhTCs, demonstrating that the integration of active non-Foster elements into time-varying media provides extraordinary ways for wave manipulation. Our theoretical framework reveals several intriguing phenomena, including the amplification of DC waves in isotropic media and an amplification rate that grows unbounded with momentum—features that lie far beyond the reach of conventional PhTCs constrained by passive material properties. These characteristics show promising applications in optical sensing and advanced devices. For instance, the ability of DC-wave amplification holds great potential for weak signal detection and signal-to-noise ratio enhancement, which is important in near-field sensing. Also, the frozen but amplified light could be helpful for an efficient optical latch or buffer. Furthermore, the linear growth of amplification with momentum provides a powerful mechanism for boosting dipole emission rates—a key to faster quantum sources—and paves the way for novel high-gain laser systems that exploit the preferential enhancement of high-$k$ modes.
While a full experimental realization presents formidable challenges, our work provides the essential theoretical foundation for this pursuit. By proposing viable implementation strategies and mapping out the stability conditions, we have laid the groundwork for future experimental efforts.


\bibliographystyle{apsrev4-2}
\bibliography{ref}

\end{document}

\clearpage


\begin{center}
    \textbf{\Large Supplementary Materials for Non-Foster PhTCs}
\end{center}
\bigskip

\setcounter{section}{0}
\setcounter{figure}{0}
\setcounter{table}{0}
\setcounter{equation}{0}

\section{Detailed Derivation of Band Structure for Time-varying Metasurface and Bulk Media}

This part, we will provide a detailed derivation of the band structure for metasurface PhTC and bulk PhTC, using transfer matrix method. In the main text we mentioned that the electric field can be expressed for each time segment within the \(n\)-th period of the time evolution as a summation of the backward and forward waves for real frequency or a summation of time-growing and time-decaying waves for purely imaginary frequency. The transfer matrix can then be obtained by using the temporal continuity conditions at the moments \((n-1)T_\mathrm{m}\) and \(nT_\mathrm{m} - t_1\). When $nT_\mathrm{m} - t_1 < t < nT_\mathrm{m}$, the electric field can be written as
\begin{equation}
    E_x = a_n e^{-\alpha_1 y} e^{j\omega_1 (t - nT_\mathrm{m})}
    + b_n e^{-\alpha_1 y} e^{-j\omega_1 (t - nT_\mathrm{m})}. \label{eq:7a}
\end{equation}
while for $(n{-}1)T_\mathrm{m} < t < (n{-}1)T_\mathrm{m} + t_2$,
\begin{equation}
    E_x = c_n e^{-\alpha_2 y} e^{j\omega_2 (t - nT_\mathrm{m})}
    + d_n e^{-\alpha_2 y} e^{-j\omega_2 (t - nT_\mathrm{m})} \label{eq:7b}
\end{equation}

Then similarly, the $y$-component of magnetic field can be derived using Faraday's law when $nT_\mathrm{m} - t_1 < t < nT_\mathrm{m}$ and $(n{-}1)T_\mathrm{m} < t < (n{-}1)T_\mathrm{m} + t_2$ respectively:
{\small
\begin{equation}
\begin{aligned}
    H_y&=\frac{\beta}{\mu_0 \omega_1}\left(a_n e^{-\alpha_1 y} e^{j\omega_1 (t-nT_\mathrm{m})}-b_n e^{-\alpha_1 y} e^{-j\omega_1 (t-nT_\mathrm{m})}\right)\\
    H_y&=\frac{\beta}{\mu_0 \omega_2}\left(c_n e^{-\alpha_2 y} e^{j\omega_2 (t-nT_\mathrm{m})}-d_n e^{-\alpha_2 y} e^{-j\omega_2 (t-nT_\mathrm{m})}\right). \label{eq.8}
\end{aligned}
\end{equation}}
Note that the spatial term $e^{-jkz}$ is omitted in Eqs.~(\ref{eq:7a}), (\ref{eq:7b}), and (\ref{eq.8}) for brevity.

At a temporal boundary, the quantities that must remain continuous before and after the jump (\(t = t_0\)) are the electric displacement field \(\mathbf{D}\) and the magnetic flux density \(\mathbf{B}\)~\cite{morgenthaler1958velocity, mendoncca2002time}. Applying this condition to our metasurface case, and noting the surrounding media is not changed (vacuum), we obtain: 
$\varepsilon_0 E_x(t = t_0^-) = \varepsilon_0 E_x(t = t_0^+), \quad \mu_0 H_y(t = t_0^-) = \mu_0 H_y(t = t_0^+).$
Note that the magnetic flux has two components $B_y$ and $B_z$. However, only \(B_y\) is continuous in time, as explained in \cite{wang2023controlling}. 

Next, for simplicity, we consider a point infinitesimally close to the metasurface, i.e., \(y \to 0^+\). Since the fields above the metasurface in vacuum must remain continuous in time, this results in two equations describing temporal continuity at the moments \((n-1)T_\mathrm{m}\) and \(nT_\mathrm{m} - t_1\). Accordingly, in the matrix form, we have: 
\begin{subequations}
\label{eq:9}
\begin{equation}
\begin{pmatrix}
1 & 1 \\
1 & -1
\end{pmatrix}
\begin{pmatrix}
a_{n-1} \\
b_{n-1}
\end{pmatrix}
=
\begin{pmatrix}
e^{-j\omega_2 T_m} & e^{j\omega_2 T_m} \\
\frac{\omega_1}{\omega_2} e^{-j\omega_2 T_m} & -\frac{\omega_1}{\omega_2} e^{j\omega_2 T_m}
\end{pmatrix}
\begin{pmatrix}
c_n \\
d_n
\end{pmatrix}
\end{equation}
\begin{equation}
\begin{pmatrix}
e^{-j\phi_1} & e^{j\phi_1} \\
e^{-j\phi_1} & -e^{j\phi_1}
\end{pmatrix}
\begin{pmatrix}
a_{n} \\
b_{n}
\end{pmatrix}
=
\begin{pmatrix}
e^{-j\phi_2} & e^{j\phi_2} \\
\frac{\omega_1}{\omega_2} e^{-j\phi_2} & -\frac{\omega_1}{\omega_2} e^{j\phi_2}
\end{pmatrix}
\begin{pmatrix}
c_n \\
d_n
\end{pmatrix} \label{eq.5b}
\end{equation}
\end{subequations}
where $\phi_1=\omega_1 t_1$ and $\phi_2=\omega_2 t_1$. Combining Eqs.~(\ref{eq:9}a) and (\ref{eq:9}b) yields the final transfer-matrix describing the relation between the input field and output field through one temporal period: $(a_{n-1},b_{n-1})^T=\mathbf{A}\cdot(a_{n},b_{n})^T$. Here, $T$ refers to the transpose operation, and $\mathbf{A}$ is a square matrix. 

Then, in order to derive the band structure, we need to admit that in an ideal infinite periodic time modulation, the amplitude of the field on adjacent temporal periods should be the same, except for a fixed phase difference which can be given by Floquet theorem, $(a_{n},b_{n})^T=e^{j\omega_{\rm F} T_m}(a_{n-1},b_{n-1})^T$, where $\omega_F$ is the Floquet frequency. Finally, we arrive at the eigenvalue problem which reads,
\begin{equation}
\Big(\mathbf{A}-e^{-j\omega_{\rm F} T_m} \mathbf{I}\Big)(a_{n},b_{n})^T
=0, 
\label{eq: eig}
\end{equation}
where $\mathbf{I}$ is an identity matrix. 

To obtain a non-trivial solution for Eq.~(\ref{eq: eig}), the determinant of the matrix within the parentheses must be zero, yielding the dispersion relation between $\omega_{\rm F}$ and $\beta$. This completes the derivation of the band diagram for a time-varying capacitive metasurface. 

Next, we show how to calculate the band structure of a bulk PhTC, although the derivation has been thoroughly discussed in~\cite{asgari2024theory}, the key steps of the derivation will be outlined in this section for the sake of readability.

The time-varying progress discussed here still refers to the periodic square-wave modulation introduced in the main text:
\begin{equation} \varepsilon(t)=
\begin{cases}
    \varepsilon_2, \quad (n-1) T_\mathrm{m}<t<(n-1) T_\mathrm{m}+t_2\\
    \varepsilon_1, \quad n T_\mathrm{m}-t_1<t<n T_\mathrm{m}
\end{cases}
, n \in \mathbb{Z}.
\end{equation}
Accordingly, the transfer matrix method and Floquet’s theorem can likewise be employed to obtain the band diagram. Similarly, let's see what happens if permittivity is a dispersionless negative value before deriving band diagram. The dispersion relation of plane wave in a dielectric bulk medium (no magnetism involved, $\mu_r=1$) is well known as: $\omega \sqrt{\varepsilon_r \varepsilon_0 \mu_0}=k$. Now let us assume the wavenumber to be a positive real value since we consider a lossless system. We notice that in this case, if $\varepsilon_r<0$ then the frequency must be a purely imaginary value as the case in negative capacitance metasurface in the main text. Consequently, the eigen mode in a negative permittivity bulk medium is a stationary wave in space but exponentially growing or decaying in time. After confirming this, we can now proceed to derive the band diagram.

The electric displacement field $D_x$ and magnetic flux $B_y$ inside the two time segments in the n-th time period ($n T_\mathrm{m}-t_1<t<n T_\mathrm{m}~and~(n-1) T_\mathrm{m}<t<(n-1) T_\mathrm{m}+t_2$) can be expressed as the summation of forward and backward wave with different frequencies, respectively.

\begin{equation}
\begin{aligned}
    D_x&=(a_n  e^{j\omega_1 (t-nT_\mathrm{m})}+b_n  e^{-j\omega_1 (t-nT_\mathrm{m})})e^{-jk z}\\
    D_x&=(c_n  e^{j\omega_2 (t-nT_\mathrm{m})}+d_n  e^{-j\omega_2 (t-nT_\mathrm{m})})e^{-jk z}\\
    B_y&=\frac{\mu_0 \omega_1}{k}(a_n  e^{j\omega_1 (t-nT_\mathrm{m})}-b_n  e^{-j\omega_1 (t-nT_\mathrm{m})})e^{-jk z}\\
    B_y&=\frac{\mu_0 \omega_2}{k}(c_n  e^{j\omega_2 (t-nT_\mathrm{m})}-d_n  e^{-j\omega_2 (t-nT_\mathrm{m})})e^{-jk z},
\end{aligned}
\end{equation}
where $a_n$, $b_n$, $c_n$ and $d_n$ denotes the complex amplitudes of each wave component. $\omega_1$ and $\omega_2$ are related to each other since the wave momentum is conserved at the temporal interface:
\begin{equation}
    \omega_1^2\varepsilon_{r1}=\omega_2^2\varepsilon_{r2}. \label{eq.s3}
\end{equation}
Then apply the continuous condition which imposes $D_x$ and $B_y$ to be continuous at the moment of $(n-1)T_\mathrm{m}$ and $n T_\mathrm{m}-t_1$:

\begin{equation}
    \begin{bmatrix} 
    1 & 1 \\ 
    1 & -1 
    \end{bmatrix} \cdot 
    \begin{bmatrix} 
    a_{n-1} \\ 
    b_{n-1} 
    \end{bmatrix} = 
    \begin{bmatrix} 
    e^{-j\omega_2 T_m} & e^{j\omega_2 T_m} \\ 
    \frac{\omega_2}{\omega_1}e^{-j\omega_2 T_m} & -\frac{\omega_2}{\omega_1}e^{j\omega_2 T_m} \end{bmatrix} \cdot 
    \begin{bmatrix} c_n \\ 
    d_n 
    \end{bmatrix}. \label{eq.4}
\end{equation}

\begin{equation}
    \begin{bmatrix} 
    e^{-j\omega_1 t_1} & e^{j\omega_1 t_1} \\ 
    \frac{\omega_1}{\omega_2}e^{-j\omega_1 t_1} & -\frac{\omega_1}{\omega_2}e^{j\omega_1 t_1} \end{bmatrix} \cdot 
    \begin{bmatrix} a_n \\ 
    b_n 
    \end{bmatrix} = 
    \begin{bmatrix} e^{-j\omega_2 t_1} & e^{j\omega_2 t_1} \\ 
    e^{-j\omega_2 t_1} & -e^{j\omega_2 t_1} 
    \end{bmatrix} \cdot 
    \begin{bmatrix} c_n \\ 
    d_n 
    \end{bmatrix}. \label{eq.5}
\end{equation}
Combining Eq.~(\ref{eq.4}) and Eq.~(\ref{eq.5}), we get the relationship of the field amplitudes of two adjacent unit cells which is also referred to as transfer matrix.
\begin{equation}
    \begin{bmatrix}
        a_{n-1}\\
        b_{n-1}
    \end{bmatrix}=
    \begin{bmatrix}
        A &B\\
        C &D
    \end{bmatrix} \cdot
    \begin{bmatrix}
        a_n\\
        b_n
    \end{bmatrix}.
\end{equation}

To obtain the band diagram, Floquet theorem is needed which requires the amplitudes of the wave to remain unchanged between two adjacent time periods except for a fixed phase change determined by Floquet frequency $\omega_F$,
\begin{equation}
    \begin{bmatrix}
        a_n\\
        b_n
    \end{bmatrix}=
    e^{j \omega_F T_\mathrm{m}}
    \begin{bmatrix}
        a_{n-1}\\
        b_{n-1}
    \end{bmatrix}.
\end{equation}
Finally an eigenvalue problem is obtained:
\begin{equation}
    \begin{bmatrix}
        A &B\\
        C &D
    \end{bmatrix} \cdot
    \begin{bmatrix}
        a_n\\
        b_n
    \end{bmatrix}=
     e^{-j \omega_F T_\mathrm{m}}
    \begin{bmatrix}
        a_n\\
        b_n
    \end{bmatrix}.
\end{equation}
Solving this eigenvalue problem, the band diagrams could be obtained as shown in the main text.


\section{Stability Considerations}

\begin{figure}
    \centering
    \includegraphics[width=1\linewidth]{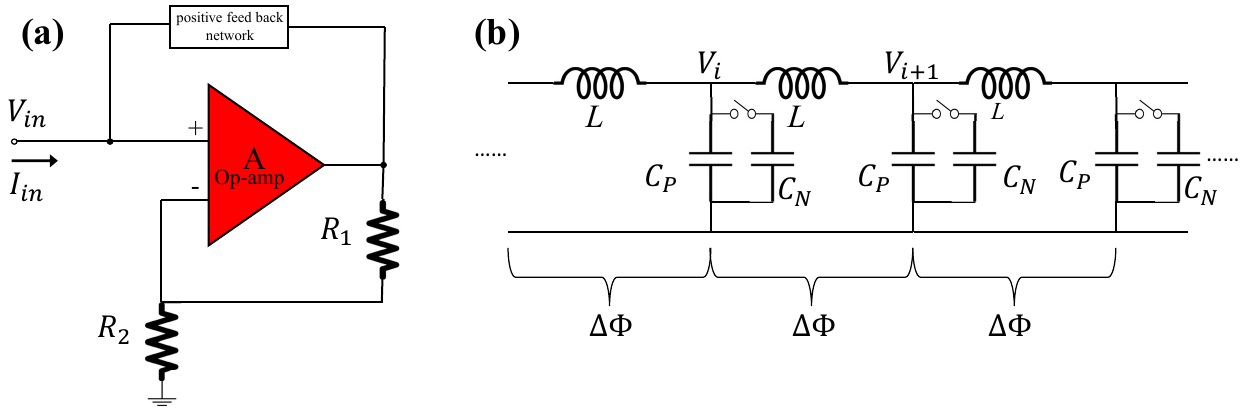}
    \caption{(a) A classic Opamp-based NIC circuit, where the Opamp is an ideal voltage amplifier with a constant gain. And (b), the circuit realization of the positive-negative capacitance-switching transmission line with the electrical length $\Delta \Phi$. $L$, $C_P$ represent the effective inductance and capacitance, respectively, while $C_N$ represents the negative capacitance generated by NIC circuit.}
    \label{fig:4}
\end{figure}

Despite the promising characteristics exhibited by PhTCs combined with non-Foster elements, the stability of the system should be considered very carfefully.  It is known that PhTCs can inherently be unstable due to the presence of momentum bandgaps. However, the main stability issue stems from the use of non-Foster elements, which are typically implemented using NIC and NIV circuits~\cite{marius_Negative_1928, merrill_Theory_1951, linvill_Transistor_1953,larky1957negative, schwarz_Negative_1969a}, as shown in Fig.~\ref{fig:4}(a). These circuits are essentially specialized amplifiers that include the load to be inverted within a positive feedback loop. The use of positive feedback assures either  impedance sign flipping in the case of NICs ({$Z_{in}\approx-Z_{load}$}) or impedance sign flipping accompanied with the impedance inversion ({$Z_{in}\approx-1/Z_{load}$}), in the case of NIVs. Here,  \textit{Z}\textsubscript{in} and \textit{Z}\textsubscript{load } stand for  the input and load impedance, respectively.  A typical Opamp-based NIC circuit  that provides negative admittance is  shown in Fig.~\ref{fig:4}(a).  A detailed explanation of operation of this circuit can be found in \cite{zanic2024stability}.  Briefly, in this configuration, the OPamp operates as a simple voltage amplifier that rises potential of the right node of the positive feedback network to $A_V V_{in}$ , while the left node is at the (original) input potential $V_{in}$.  Thus, there is a superposition of  input current ($I_{in}$ ) and the current supplied from the output of the OPamp, which builds up a voltage drop across the feedback network. A simple calculation \cite{zanic2024stability} shows that input admittance becomes a  scaled 'negative image' of the admittance of feedback network. It is important to point out that this discussion is only valid when the OPamp is operating in the linear range. If the input signal is high enough, the OPamp is driven into the non-linear region, and admittance and impedance are no longer a simple inversion of each other ($Y\neq 1/Z$). Therefore, both in NICs and NIVs, the impedance and admittance should be treated separately \cite{schwarz_Negative_1969a}.  Furthermore, the positive feedback loop in an Opamp-based NIC circuit can make this circuit unstable.  Suppose, for example, that the input of the circuit in Fig.~\ref{fig:4}(a) is left open. Ideally, both the input and output voltages are zero, and the circuit should be stable. In reality, however, there is always some noise at the input, which is amplified and appears at the output of the amplifier. This amplified noise is fed back into the input, then amplified again, fed back into the input, and so on... This process naturally leads to instability. In practice, the NIC circuit is never used 'alone', but it is connected to some external passive network, and one should analyze the stability of the whole system. It is done by assuming the presence of some 'disturbing signal' and investigating the response of the system in the Laplace domain \cite{zanic2024stability, hrabar2018first, hrabar2023accurately}.

Here, we analyze a bulk medium with positive-negative switched permittivity as an example to briefly illustrate how to determine stability. It is known that an unbounded homogeneous and isotropic bulk medium can be equivalently represented as a uniform transmission line. On the other hand, the homogeneous transmission line can be modeled by a $N$-cell LC ladder network in the limit of $\Delta \Phi < \pi/2$, where $\Delta \Phi$ is the electrical length~\cite{caloz2005electromagnetic}. Thus, bulk material with positive-negative modulated permittivity can be mimicked by periodically connecting or disconnecting switches connected to a non-Foster capacitor, as illustrated in Fig.~\ref{fig:4}(b). 

Now we focus on the stability of the non-Foster-capacitor-loaded $N$-cell LC ladder network. By transforming the real circuit component into Laplace domain component, we derive the transfer function which is defined as the output/input ratio of the voltage at the port of each unit cell in the following form: $H_i(s)=\frac{V_{i+1}(s)}{V_i(s)}=\frac{1}{1+s^2L(C_P+C_N)}$, where $L$, $C_P$ and $C_N$ represent the equivalent inductance, capacitance and the negative capacitance created by NIC circuit. Since the transmission line is homogeneous, the transfer function of each unit cell is the same, and the final transfer function of the transmission line can be expressed as $H^N(s)$. Using Nyquist method as discussed in detail in~\cite{hrabar2023accurately} and~\cite{zanic2024does}, the stability condition can be determined by the denominator of $H^N(s)$. Thus, in our lossless non-Foster loaded LC network, the only possible way that it becomes stable is $C_P>\left | C_N\right |$, which requires the total capacitance to be larger than zero and consequently, the effective permittivity must also be greater than zero \cite{Rengarajan_2013, loncar2016stability}. However, the above conclusion is based on the assumption of an ideal negative capacitance and the system is lossless, meaning that the negative capacitance obtained from the NIC circuit is non-dispersive with infinite bandwidth, and the unstable positive real DC poles always exist~\cite{hrabar2018improving}. These unstable DC poles can exist when the Opamp work at $\omega=0$, that is, an ideal Opamp with infinite bandwidth or a low-pass Opamp, and this is why our non-Foster PhTCs can amplify DC waves. Indeed, ideally dispersionless non-Foster elements violate causality  and cannot be realized~\cite{okorn2011investigation, tretyakov2007veselago}. Therefore, it is physically incorrect to analyze the stability condition under the assumption of an ideal dispersionless non-Foster element. Importantly, the stability condition is different if we consider realistic elements. 

Recently, it was proposed that a low-pass and band-pass negative capacitance can be achieved with realistic circuits~\cite{hrabar2019investigation} \cite{loncar2016stability} as well as switching between positive and negative capacitance~\cite{hrabar2020stable} \cite{zanic2024stability}. The two different realistic negative capacitance models have distinct stability conditions. In fact, the low-pass negative capacitance provides an external stability condition compared to the ideal one, i.e., enabling the total capacitance or the permittivity to be negative. However, the low-pass characteristic can not remove the unstable DC poles. Consequently, DC wave amplification can still be achieved but with a little improved stability condition. While the band-pass negative capacitance eliminates the presence of the unstable DC poles, thereby providing better stability robustness for the entire network. Paradoxically, amplifying DC waves would no longer be possible in this case.

In conclusion, this section has examined the challenge of realizing an active system and has presented one possible strategy for ensuring stability, including using low-pass or band-pass negative capacitor. Both of them offer an elegant solution in which stability is achieved by design. The use of low-pass or band-pass negative capacitor is based on the choice of DC wave amplification or more robust stability properties. Although both the low-pass and band-pass models are dispersive, a well-designed low-pass and band-pass capacitor can exhibit nearly non-dispersive behavior within its operational bandwidth. This ensures the validity of our theory used in this Letter and potentially preserves the exotic, near-linear growth of the amplification rate.

\section{Harmonic Distribution Calculation}

Calculating the harmonic distribution is not such straightforward when using transfer-matrix method to calculate band structure. So, it is necessary to introduce the technique we use for calculating the harmonic distribution. First, let's review the properties of the band structure. One can see from the eigenvalue problem Eq.~(\ref{eq: eig}) that, for a given wavenumber, there will always be an infinite series of frequencies $\omega_n=\omega_{\mathrm{F}}+n\omega_{\mathrm{m}}$. Our goal is to find the amplitudes of all signal components associated with each $\omega_n$. We first arbitrarily choose one $\omega_n$ such as the fundamental frequency ($n=0$), and then calculate the corresponding eigenvector $\left[a_n,b_n\right]^{\mathrm{T}}$, which are amplitudes of fields in one time segment of one modulation period $T_{\mathrm{m}}$. Next, by using Eq.~(\ref{eq.5b}) the amplitudes corresponding to the fields in another time segment of the same modulation period, i.e., $\left[c_n,d_n\right]^{\mathrm{T}}$ can be obtained. Now we have the total field in one modulation period in time domain. But one needs to be careful that for the fundamental frequency that is at half of the modulation frequency $\omega_{\mathrm{m}}/2$, the same steps need to be repeated to obtain the total field in the next modulation period, i.e., $\left[a_{n+1},b_{n+1}\right]^{\mathrm{T}}$ and $\left[c_{n+1},d_{n+1}\right]^{\mathrm{T}}$, since one time period for this fundamental frequency is $2 T_{\mathrm{m}}$. Then we only need to normalize the total field by dividing by \( e^{\Im(\omega_{\rm F}) t} \) to ensure a periodic function and then perform the Fourier transform on this normalized time-domain signal, the final harmonic distribution can be obtained.


\end{document}